\documentclass[10pt,conference]{IEEEtran} 
\IEEEoverridecommandlockouts
\usepackage{cite}
\usepackage{pgfplots}
\usepackage{pgfplotstable}
\pgfplotsset{compat=1.18}
\usepackage{booktabs}
\usepackage{url}
\usepackage{amsmath,amssymb,amsfonts}
\usepackage{algorithmic}
\usepackage{graphicx}
\usepackage{textcomp}
\usepackage{xcolor}
\usepackage{balance}
\def\BibTeX{{\rm B\kern-.05em{\sc i\kern-.025em b}\kern-.08em
    T\kern-.1667em\lower.7ex\hbox{E}\kern-.125emX}}

\usepackage[breakable]{tcolorbox}
\tcbset{breakable}
    
\usepackage{xcolor}
\usepackage{xspace}

\PackageWarning{TODO:}{X}
\PackageWarning{TODO:}{X\%}

\usepackage{tcolorbox}

\definecolor{custom-blue}{rgb}{0,0,1}

\begin{document}

\title{Can Perplexity Serve as a Cognitive Signal for Code Understandability?}

\author{
\IEEEauthorblockN{
Xiaokai Rong\textsuperscript{$\dagger$},
Mohammadali Sefidi Esfahani\textsuperscript{$\ddagger$},
Aashish Yadavally\textsuperscript{$\S$},
Rigby Peter\textsuperscript{$\ddagger$}, and
Tien N. Nguyen\textsuperscript{$\dagger$}}
\IEEEauthorblockA{
xiaokai.rong@utdallas.edu, mohammadali.sefidi@gmail.com,\\
aashish.yadavally@ucf.edu, peter.rigby@concordia.ca,
Tien.n.nguyen@utdallas.edu}
\thanks{\textsuperscript{$\dagger$}The University of Texas at Dallas;
\textsuperscript{$\ddagger$}Concordia University;
\textsuperscript{$\S$}University of Central Florida.}
}

\maketitle

\begin{abstract}
Recent work suggests that token-level perplexity from large language models can align with localized human confusion during code comprehension. This raises a natural question: can perplexity also serve as a snippet-level signal for code understandability? We conduct an empirical study of this question across multiple human-grounded datasets, including method-level understandability judgments, output-prediction tasks, and accepted understandability-improvement patches. 
Despite prior token-level evidence, we find that simple snippet-level aggregations of token perplexity, such as average, median, or peak perplexity, do not reliably correlate with human understandability. We then investigate why this happens. First, token perplexity is highly skewed and heavy-tailed across code structures; extreme spikes arise not only from semantically meaningful constructs, but also from identifiers, literals, types, separators, and tokenization artifacts. Second, human understandability labels often lack consensus, making whole-snippet difficulty a noisy target. Third, perplexity distributions and their alignment with human difficulty vary substantially across models and tokenizers. These findings explain why prior token-level perplexity--confusion alignment does not directly transfer to snippet-level understandability. Overall, our study positions perplexity as a promising but delicate cognitive signal: useful for localized code confusion, but requiring code-aware aggregation, consensus-aware evaluation, and model-sensitivity analysis before it can support reliable code-understandability measurement.
\end{abstract}


\section{Introduction}


Developers spend a substantial portion of their time reading and understanding source code. Consequently, measuring how difficult a code snippet is to understand has been an important problem. Accurate understandability measures can help developers identify hard-to-maintain code, guide refactoring, support code review, and estimate the effort required for software maintenance tasks. Traditionally, researchers and practitioners have relied on code readability and code complexity metrics as proxies for this goal. Readability metrics capture surface-level properties such as identifier names, formatting, comments, indentation, and visual structure. Complexity metrics, such  as cyclomatic complexity~\cite{mccabe1976complexity} and Halstead measures~\cite{halstead1977elements}, capture structural properties including branching, nesting, control-flow paths, and token-level counts. However, prior empirical work has shown that these metrics are not reliable indicators of whether developers actually understand a code snippet~\cite{scalabrino2021automatically}. A snippet may look readable but still be difficult to understand because of unfamiliar APIs, hidden semantic dependencies, misleading names, or non-obvious behavior. Conversely, structurally complex code may still be understandable to developers who recognize its pattern or domain intent.

Recent work suggests an alternative direction: using model uncertainty as a cognitive signal. In particular, token-level perplexity (PPL) from large language models (LLMs) measures how unexpected a code token is given its preceding context. Abdelsalam {\em et al.} \cite{abdelsalam2025humansllmsprocessconfusing} showed that token-level perplexity can align with localized human confusion signals measured through eye tracking and EEG fixation-related potentials. This finding suggests that LLM surprisal may capture aspects of human cognitive processing during code comprehension. If such localized {\bf \em token-level} signals can be aggregated appropriately, they may provide a new way to estimate {\bf \em snippet-level} code difficulty and understandability.


However, moving from token-level perplexity as a signal of localized confusion to snippet-level perplexity as a signal of overall code difficulty is non-trivial. Human confusion is often localized: a developer may be confused by a particular operator, predicate, API, literal, or expression. Code understandability, in contrast, is typically measured at the level of a snippet or a full method through behavioral outcomes such as comprehension-question accuracy or execution-level correctness. A direct~aggregation of token perplexities, such as taking the mean,~median, or maximum over all tokens, may lose the cognitive meaning of the signal. 


To investigate this, we first conducted a study~across the available snippet-level and pairwise evidence sources: the Understandability dataset introduced by Scalabrino {\em et al.}~\cite{scalabrino2021automatically}, the UTD output-prediction dataset~\cite{nguyen2026effectcodeobfuscationhuman}, and the CodeUP accepted-change dataset~\cite{oliveira2025understanding}. We compute token-level perplexity for each code unit using an LLM and aggregate these values using simple snippet-level or pairwise statistics such as sequence-level and peak/tail perplexity. We then compare these aggregated scores against the corresponding human-grounded target for each dataset. The preliminary results show {\bf \em no reliable correlation} between simple perplexity aggregation and human understandability or accepted understandability-improvement direction. This result motivates a deeper empirical investigation into {\bf \em why perplexity fails at the snippet level} and {\bf \em what factors must be addressed} before perplexity can serve as a cognitively grounded code-understandability signal.

Our study examines three key challenges. \underline{First}, we study the {\bf \em distribution of token-level perplexity inside code snippets}.~We find that perplexity is highly skewed and heavy-tailed across code structures. Extreme spikes occur not only in semantically meaningful constructs such as predicates, loops, and branch statements, but also in less informative categories such as type names, literals, and separators. This makes naive aggregation unstable, because a small number of extreme tokens can dominate a snippet-level score. This suggests that any future snippet-level metric must aggregate perplexity
values for a code snippet over cognitively meaningful regions, such as statements, predicates, expressions, method calls, control/data-flow slices, rather than treating it as a flat token sequence.

\underline{Second}, we study the role of {\bf \em human-label consensus}.~Human understandability labels are inherently noisy: the same snippet may be easy for one developer but difficult for another, depending on experience, domain knowledge, and reasoning strategy \cite{scalabrino2021automatically,peitek2020readingorder}. We observe that many snippets fall into a mixed-response band where participants do not clearly agree if the code is easy or hard. This ambiguity weakens the apparent alignment between perplexity and human understandability. Consensus filtering makes the human-difficulty target more coherent, but also reduces the available sample size. This highlights the need for a human study to build an understandability dataset suitable for the study of the relationship between perplexity and human understandability at the snippet level.

\underline{Third}, we study {\bf \em model dependence}. Perplexity is not an intrinsic property of code alone; it depends on the model and tokenizer used to compute token probabilities. Our experiments show that perplexity distributions and perplexity--understandability correlations vary substantially across models. Some models show strong alignments with human understandability in our setting, while others show weak or~even negative correlations. This result indicates that perplexity-based complexity measures must consider model sensitivity and should not be assumed to generalize across LLMs.


This paper makes the following contributions:

1. We empirically show that naive snippet-level perplexity aggregation fails to align with human understandability.

2. We identify three causes: heavy-tailed perplexity distributions, human-label disagreement, and model dependence.

3. We compare several human-grounded evidence types, including whole-snippet behavioral labels, output prediction, patch-level improvement, and localized FRP-AoC evidence.

4. We report the implications for designing future model-based understandability metrics.

\section{Background and Definitions}
\label{sec:background}


{\bf \em Code readability} refers to how easily the surface form of code can be read and visually processed by humans~\cite{buse2010learning}. Readability is influenced by lexical, formatting, and presentation-level properties, e.g., identifier names, comments, indentation, line length, syntax highlighting, alignment, and other visual/spatial features. Prior readability models~\cite{posnett2011simple,buse2010learning,daka2015modeling,scalabrino2016improving,scalabrino2018comprehensive} often rely on structural and textual features of the code and evaluate them against developers' perceived readability judgments.
{\bf \em Code complexity}~\cite{mccabe1976complexity} refers to structural or cognitive properties that may affect comprehension effort, e.g., the number of independent execution paths, nesting depth, control structures, size, coupling, or other source-level properties. Metrics such as cyclomatic complexity~\cite{mccabe1976complexity}, Halstead-style metrics~\cite{halstead1977elements}, nesting depth, number of statements, and related cognitive-complexity measures are often used as proxies for reasoning burden. 
These metrics capture important structural aspects of code.


{\bf \em Code understandability}~\cite{scalabrino2021automatically} is broader than readability and complexity. Readable or structurally simple code may still be difficult to understand if the developer cannot infer its purpose, the relationships among code entities, or the latent semantics behind API calls and program behavior. Code understandability~\cite{scalabrino2021automatically,oliveira2025understanding}, therefore, concerns the extent to which developers can build a correct mental model of what the code does, including low-level elements such as statements, beacons, and motifs, as well as higher-level structures such as classes, packages, and abstractions. Unlike readability and complexity, understandability depends not only on the code's surface form and structure, but also on the developer's prior knowledge, domain familiarity, API knowledge, and mental models. Scalabrino {\em et al.}~\cite{scalabrino2021automatically} emphasize that readability and complexity proxies should not be treated as equivalent to understandability: they may capture aspects related to comprehension, but {\em they do not directly measure whether a developer actually understands a code snippet}.

{\bf \em Human confusion/difficulty}~\cite{abdelsalam2025humansllmsprocessconfusing} refers to localized difficulty experienced by developers when a code region violates their expectations or disrupts their mental model during comprehension. Abdelsalam {\em et al.}~\cite{abdelsalam2025humansllmsprocessconfusing} measure human confusion~using EEG-based fixation-related potentials, focusing on late frontal positivity, a neurophysiological response associated with unexpected but plausible input during comprehension. Human confusion and code understandability are related but distinct. While understandability describes if a developer can form a correct mental model, human confusion captures cognitive disruptions that occur at specific tokens, expressions, or regions.

{\bf \em Perplexity}~\cite{miaschi-etal-2021-makes} and token surprisal quantify how unexpected a code token is under a pretrained code model. Importantly, Abdelsalam {\em et al.}~\cite{abdelsalam2025humansllmsprocessconfusing} showed that token-level perplexity correlates with human surprise and confusion signals measured through eye tracking and EEG fixation-related potentials during code comprehension. This finding suggests that model uncertainty may capture aspects of human cognitive processing. 

Abdelsalam {\em et al.}~\cite{abdelsalam2025humansllmsprocessconfusing} provide evidence that token-level perplexity aligns with localized human confusion at specific tokens or AOI-level regions. Our work asks {\bf \em whether this localized signal can be lifted to snippet-level code understandability}, where human labels are behavioral, code units are larger, and aggregation across many tokens is required.

\section{Code Understandability Datasets}
\label{sec:method}

In our empirical study, we used the following datasets.

\paragraph{Whole-method behavioral understandability dataset ($\alpha$ dataset~\cite{scalabrino2021automatically})}
We use the code-understandability dataset~\cite{scalabrino2021automatically}, which was designed to study whether code-, documentation-, and developer-related metrics correlate with human code understandability. The dataset contains 50 Java/Android methods selected from 10 open-source systems, with methods chosen to be non-trivial but still suitable for human comprehension tasks. In the original study, 63 Java developers and computer science students each inspected up to eight randomly assigned methods, resulting in 428 human evaluations. For each method, participants first indicated whether they perceived that they understood the method, and the system recorded the time spent before this decision. If participants reported that they understood the method, they then answered three verification questions about the code, which were used to measure their actual understanding. Thus, the dataset provides several human-grounded understandability proxies: Perceived Binary Understandability, Time Needed for Perceived Understandability, Actual Understandability, and Timed Actual Understandability. In our study, these proxies serve as ground-truth signals for evaluating whether perplexity-based aggregation aligns with human code comprehension difficulty.

\paragraph{Secondary behavioral dataset using Output Prediction ($\beta$ dataset~\cite{nguyen2026effectcodeobfuscationhuman})}
We use the code comprehension dataset~\cite{nguyen2026effectcodeobfuscationhuman}, which studies how humans understand Python and JavaScript programs via output-prediction tasks. In the study, 50 undergraduate CS students at 
University of Texas at Dallas~(UTD) were asked to compute the exact output of a given function and input, while response correctness, response time, and self-reported programming experience were recorded. In~our study, we use only the unobfuscated code  and take output-computed accuracy as human-grounded understandability signals: code with lower accuracy are treated as more difficult/confusing.

\paragraph{Patch-level Transfer/Stress Test ($\gamma$ dataset~\cite{oliveira2025understanding})}
We use the CodeUP dataset~\cite{oliveira2025understanding} containing code review comments and corresponding code changes aimed at improving code understandability. The original study manually analyzed Java pull-request comments from open-source GitHub projects and identified reviewer suggestions related to understandability issues.
For our analysis, we focus on accepted understandability-improvement pairs, where the reviewer’s suggestion was integrated into the codebase. We treat each pair as a relative proxy: the before-change version is considered less understandable, while the after-change one is considered more understandable. 

\begin{table}[t]
\centering
\scriptsize
\caption{Datasets and roles. The aliases denote Understandability ($\alpha$~\cite{scalabrino2021automatically}), UTD output prediction ($\beta$~\cite{nguyen2026effectcodeobfuscationhuman}), and CodeUP ($\gamma$~\cite{oliveira2025understanding}). }
\label{tab:dataset-roles}
\tabcolsep 2.4pt
\vspace{-6pt}
\begin{tabular}{llrcc}
\toprule
Dataset & Understandability Proxies & Items & \#Responses & \#persons  \\
\midrule
$\alpha$~\cite{scalabrino2021automatically} & question-answer verification correctness & 50 & 428 & 63  \\
$\beta$~\cite{nguyen2026effectcodeobfuscationhuman}  & output-prediction correctness & 20 & 121 & 48   \\
$\gamma$~\cite{oliveira2025understanding} & code review understandability improvement & 171 & -- & --  \\
FRP-AoC~\cite{abdelsalam2025humansllmsprocessconfusing} & AOI-level neurophysiological  signal & 144 & 1,727 & 24  \\
\bottomrule
\end{tabular}
\end{table}

\paragraph{Dataset Roles}

Table~\ref{tab:dataset-roles} summarizes the role of each dataset. The $\alpha$~\cite{scalabrino2021automatically} and $\beta$~\cite{nguyen2026effectcodeobfuscationhuman} datasets provide behavioral human-difficulty labels. 
The $\gamma$ dataset~\cite{oliveira2025understanding} provides accepted understandability-improvement patches and is used only as an external stress test.
This distinction is important since the datasets do not provide the same kind of ground truth. The $\alpha$~\cite{scalabrino2021automatically} and $\beta$~\cite{nguyen2026effectcodeobfuscationhuman} datasets ask whether perplexity aligns with whole-snippet behavioral difficulty. The $\gamma$ dataset~\cite{oliveira2025understanding} asks whether a score decreases after accepted understandability improvements. We interpret the results within each dataset's evidence type rather than treating all datasets as equally.

\section{Empirical Results}
\label{sec:results}
Prior work shows that traditional readability and complexity metrics are weak proxies for code understandability~\cite{scalabrino2021automatically}. Recently, token-level LLM perplexity has been shown to align with localized human confusion signals measured through eye-tracking and EEG fixation-related potentials~\cite{abdelsalam2025humansllmsprocessconfusing}. These findings raise a natural question as we ask in RQ1.

\noindent {\bf RQ1: [From Token-Level Perplexity to Snippet-Level Understandability]} Can the localized token-level PPL signal be lifted to snippet-level code understandability?

From the results for RQ1 (Section~\ref{sec:prelim}), we found that 
simple snippet-level aggregations of token perplexity do not provide a reliable proxy for human code understandability. This motivates us to further investigate how perplexity behaves at the snippet level and what factors must be addressed before perplexity can serve as a cognitively grounded signal for code understandability. Thus, we seek to answer the questions:

\noindent {\bf RQ2: [Perplexity Distribution]} How is token-level perplexity distributed across code structures within code snippets?

\noindent  {\bf RQ3: [Human Label Noise]} How do human-label reliability and code-aware aggregation affect perplexity alignment?

\noindent  {\bf RQ4: [Model Dependence]} How do perplexity distributions depend on models?

\subsection{From Token-Level Perplexity to Snippet-Level Understandability (RQ1)}
\label{sec:prelim}


To answer RQ1, as a first-order test, we evaluate if simple aggregations of token-level perplexity already align with human behavioral understandability labels. This setting differs from localized AOI-level analysis in~\cite{abdelsalam2025humansllmsprocessconfusing}: the model assigns uncertainty across all tokens in a method, while the human label summarizes behavioral understanding of the entire snippet.

\paragraph{Dataset}
Table~\ref{tab:prelim-perplexity-alignment} uses datasets that provide a snippet-level or pairwise behavioral comparison. The $\alpha$ dataset contains 50 Java/Android methods evaluated by human participants through code comprehension tasks; the $\beta$ dataset contains 20 JavaScript/Python functions for output-prediction tasks; and the $\gamma$ CodeUP row uses before/after method pairs from accepted understandability-improvement changes. 

\paragraph{Perplexity}
We extract PPLs using Qwen2.5-Coder-7B under teacher forcing. The model is not asked to generate code; instead, it assigns probabilities to the observed next tokens in the original code. From these probabilities, we compute token-level surprisal and perplexity values. We consider two simple snippet-level baselines: sequence-level perplexity value and perplexity peak value. Sequence-level perplexity summarizes the overall predictability of the method, while perplexity peak captures the largest localized token-level spike in the method.


\paragraph{Human Understandability Labels}
For the $\alpha$ dataset, we use Actual Understandability (AU) and Actual Binary Understandability (ABU) as behavioral proxies for code understandability. 
For AU, human participants were asked a set of $q$ verification questions about each snippet. AU captures the proportion of correct answers provided by the participants. 
This continuous score represents a degree of understanding, capturing beyond binary perception.
Formally, given a program $P \in \mathcal{D}$, let $\{a^*_1, a^*_2, \dots, a^*_q\}$ denote the gold answers and $\{a^h_1, a^h_2, \dots, a^h_q\}$ be the participants' responses. Then,
\begin{equation*}
\text{AU}(P) = \frac{1}{q}\sum_{i=1}^q \mathbf{1}\{a^h_i = a^*_i\}
\end{equation*}
For Actual Binary Understandability, ABU$_{k\%}$ is defined as a binarized variant of AU obtained by applying a threshold of $k\%$ over AU. For actual understandability score $\text{AU}(P)$:
\[
\text{ABU}(P) = 
\begin{cases}
1, & \text{if } \text{AU}(P) \geq k\%, \\
0, & \text{otherwise}.
\end{cases}
\]
Because each snippet is evaluated by multiple participants, we aggregate valid participant responses to obtain a snippet-level human label. We define behavioral difficulty as the complement of the corresponding understandability score, so larger values indicate lower human understanding. For the $\beta$ dataset, behavioral difficulty is the complement of output-prediction correctness. For the $\gamma$ dataset, we use the accepted-change direction as a pairwise comparison and test whether the before-method score is larger than the after-method score.


\begin{table}[t]
\centering
\caption{Diagnostic for naive perplexity summaries across the available whole-snippet and change datasets. For $\alpha$ and $\beta$, cells report Spearman $\rho$ with $p$ in parentheses against behavioral difficulty. For $\gamma$ CodeUP, cells report the before-method $>$ after-method win rate with one-sided sign-test $p$ in parentheses.}
\label{tab:prelim-perplexity-alignment}
\vspace{-6pt}
\scriptsize
\begin{tabular}{llrcc}
\toprule
Data & Target/comparison & $n$ & perplexity & perplexity peak/tail \\
\midrule
$\alpha$ & AU difficulty & 50 & 0.183 (.204) & 0.234 (.102) \\
$\alpha$ & ABU$_{50}$ difficulty & 50 & 0.133 (.359) & 0.142 (.326) \\
$\beta$ & output-task difficulty & 20 & 0.259 (.270) & 0.325 (.163) \\
$\gamma$ & code before $>$ code after & 171 & 0.474 (.755) & 0.439 (.879) \\
\bottomrule
\end{tabular}
\end{table}

\paragraph{Correlation Results}
Table~\ref{tab:prelim-perplexity-alignment} shows that neither sequence-level perplexity nor perplexity peak/tail provides a reliable snippet-level or pairwise proxy across the available evidence types. The observed $\alpha$ and $\beta$ correlations are weak and statistically insignificant, and the $\gamma$ before-after win rates are close to chance. We therefore interpret this preliminary result as {\bf \em lack of reliable evidence for simple snippet-level alignment} rather than proof that the true association is exactly zero. The peak/tail score is especially informative: even preserving the largest localized high-perplexity spike does not recover a reliable behavioral or accepted-change signal.

\begin{tcolorbox}[breakable=true]
\noindent\textbf{Issue 1: Token-level perplexity does not directly transfer to snippet-level understandability.}
Although prior work shows that token-level PPL can align with localized human confusion, our results show that simple snippet-level summaries such as sequence-level PPL and peak/tail PPL do not provide a reliable proxy for whole-snippet human understandability or accepted understandability-improvement direction.   \end{tcolorbox}

\begin{figure*}[t]
    \centering
    \includegraphics[width=0.98\textwidth]{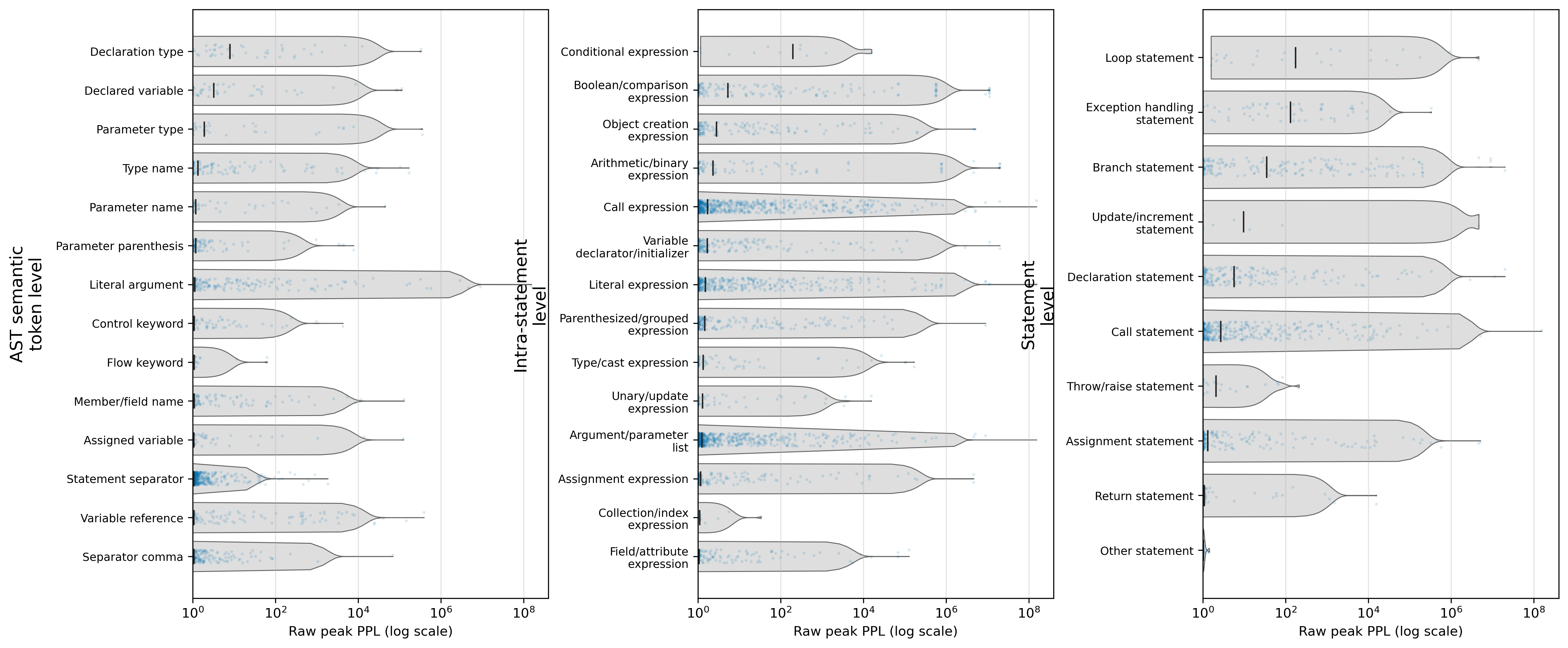}
    \vspace{-3pt}
    \caption{Perplexity distribution over the Understandability ($\alpha$~\cite{scalabrino2021automatically}) dataset at three AST-derived granularities. The within-category spread shows that PPL is heavy-tailed even after semantic AST grouping. (RQ2)}
    \label{fig:rq1-ast-distribution}
\end{figure*}

\subsection{RQ2: How is token-level PPL distributed across code structures within code snippets?}






RQ2 examines whether token-level PPL is statistically stable enough to serve as the input to snippet-level aggregation. If PPL is dominated by rare extreme values, then simple aggregation may reflect incidental lexical or tokenizer artifacts rather than human-relevant code difficulty.

\subsubsection{Empirical Settings}
In RQ2, we use the $\alpha$ dataset~\cite{scalabrino2021automatically} with the source code of 50 methods as the code-snippet corpus. We score each method with Qwen2.5-Coder-7B-Instruct under teacher forcing, so the model assigns probabilities to the observed next tokens in the original code rather than generating new code. Token-level PPL is computed from these next-token probabilities and aligned back to source-code spans.

We form AST-derived occurrences from byte spans in the target code. 
For the semantic-token view, we enumerate terminal AST tokens in a method.
We exclude whitespaces, comments, and metadata tokens, but retain executable code and code-adjacent AST terminals.
We keep these syntactic terminals in RQ2 since the diagnostic asks whether perplexity spikes remain even after each source token is assigned a concrete code role. 
For coarser views, we group AST nodes into intra-statement expression regions and statement-level~regions. 


To connect model uncertainty to code structure, we map scored source spans to AST-derived categories at three granularities: semantic token roles, intra-statement expression regions, and statement-level regions. A model token is aligned to an AST-derived occurrence when its tokenizer offset span overlaps the occurrence byte span. If multiple model tokens overlap the same occurrence, we use the largest positive aligned token PPL as that occurrence's peak PPL. This choice intentionally preserves localized upper-tail spikes because RQ2 studies whether such spikes can dominate snippet-level aggregation. Occurrences with no aligned positive model-token perplexity are omitted from the plotted distributions. We analyze PPL values and show those values on the $\log_{10}$ scale to make orders-of-magnitude variation visible and to avoid plots being dominated entirely by a few extreme values.

\subsubsection{Empirical Results}
Fig.~\ref{fig:rq1-ast-distribution} shows that PPL is structured by code category, but the structure is not stable enough for direct aggregation. Across semantic-token, intra-statement, and statement-level views, most categories have a dense low-PPL body and a long upper tail. That is, the typical token or code occurrence in a category may be easy for the model to predict, while some occurrences in the same category can be orders of magnitude larger. The trend persists from individual semantic token roles to expression regions and entire statements, so the instability is not confined to one structural level. We also repeated the diagnostic using median aligned-token PPL per occurrence; the qualitative heavy-tail pattern remains.

The heavy tail is not confined to categories that are obviously cognitively central. Some high values appear in semantically important regions, e.g., conditional expressions, boolean/comparison expressions, loops, and branch statements. However, large values also appear in categories such as type names, literals, declaration keywords, separators, and other syntactic or lexical regions. Thus, high PPL cannot be interpreted mechanically as high human difficulty. A large spike may indicate a genuinely surprising program region, but it may also come from a rare identifier, an uncommon literal, an API-specific type, or tokenizer-specific fragmentation.

This creates a major challenge for using PPL as a snippet-level complexity or understandability signal.

This creates three challenges for snippet-level aggregation. 

i) {\bf \em Naive aggregation becomes unstable and size-sensitive}. Direct aggregation is unstable and size-sensitive: a few outlier tokens can dominate the score, and longer snippets have more opportunities to contain such outliers. 

ii) {\bf \em High perplexity does not always mean high human difficulty}: it may reflect a genuinely surprising program relation, but it may also reflect lexical rarity, API specificity, or tokenization artifacts. 

iii) {\bf \em Category-level reasoning is unreliable with perplexity alone}: even within the same semantic or statement category, the spread is large, so category membership alone cannot determine whether surprise is cognitively meaningful.

An interesting question is whether the heavy tail is a code property or a property of the chosen model/tokenizer. We treat this as a perplexity value diagnostic for the selected model.
RQ2 fixes the model to isolate the distributional behavior of perplexity within code structures; RQ4 later tests whether these observations are stable across models/tokenizers.


\begin{figure*}[htbp]
    \centering
    \includegraphics[width=0.91\textwidth]{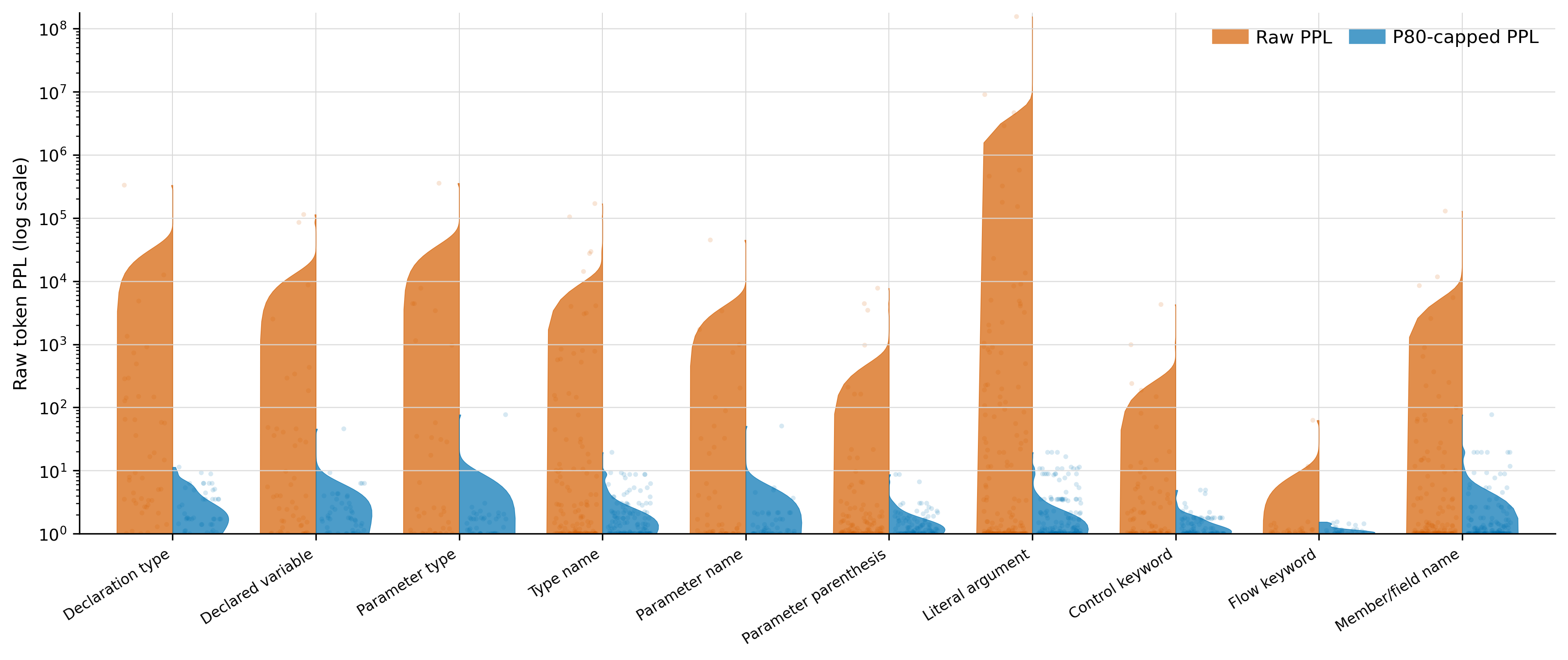}
    \vspace{-15pt}
\caption{Unfiltered and P80-capped token-level perplexity distributions over the Understandability ($\alpha$~\cite{scalabrino2021automatically}) dataset by semantic token family. The split-violin comparison shows that within-method P80 capping compresses extreme upper-tail spikes while preserving the low-perplexity distribution body. (RQ2)}  \label{fig:rq1-tail-mitigation}
\end{figure*}





\subsubsection{Top-percentile Cutoff}

We next use percentile capping as a diagnostic intervention, not as a proposed metric, to estimate how much of the perplexity distribution is driven by extreme upper-tail values.
For a percentile cap $P_k$, we compute the $k$th percentile of token perplexity within each method and replace any larger token value with that method-specific threshold. We compute capped variants for 70-90 percentile cut-offs and compare them with the uncapped token distribution using the same semantic-family grouping.

Fig.~\ref{fig:rq1-tail-mitigation} shows the results of top-percentile cutoff as a diagnostic for spike domination using the current split-violin design. For each semantic token family, the uncapped distribution and the P80-capped distribution are shown as two sides of the same violin on a log-scaled y-axis. On the unfiltered side, several families, including variable, call/member, literal, type, and declaration-keyword tokens, contain very high upper-tail observations.~These spikes indicate that token PPL is often dominated by rare or unexpected lexical events, many of which may reflect lexical rarity.~On the capped side, the P80 cap replaces token PPL values above the 80th percentile of the same snippet with that snippet-specific threshold, making the semantic-family distributions much more compact while preserving the side-by-side category comparison. To quantify this effect without tying the definition to one fixed percentile, we compute the following semantic-family upper-tail spread:
\[
\Delta_{\mathrm{tail}}^{(q)}(g)=Q_q(\log_{10}\mathrm{perplexity}_g)-\mathrm{median}(\log_{10}\mathrm{perplexity}_g),
\]
where $g$ is a semantic token family and $Q_q$ is the $q$th percentile of that family's log-scaled token perplexity values. In the reported analysis, we set $q=95$ and report the median $\Delta_{\mathrm{tail}}^{(95)}$ across semantic families for each uncapped or capped variant. The statistic confirms the visual pattern in Fig.~\ref{fig:rq1-tail-mitigation}: the median tail spread drops from 1.876 without capping to 0.567 under the P80 cap. We then repeat the same calculation for nearby top-percentile cutoffs. A looser P90 cap reduces the spread to 1.112, while a stricter P70 cap reduces it to 0.286. Thus, the conclusion is not tied to one cutoff: stronger caps compress the upper tail more aggressively, but all tested caps show that token perplexity is dominated by a relatively small number of extreme upper-tail values rather than by a uniformly shifted distribution across all tokens.

However, the comparison also explains why top-percentile cutoff is only a diagnostic and stabilization step. Percentile capping suppresses extreme spikes and makes the distributions numerically more compact, but it also changes the signal being aggregated. The cap is applied uniformly across semantic families and does not decide whether a particular high-perplexity token is a meaningful cognitive signal or an artifact of lexical rarity, formatting, API specificity, or tokenization. Thus, percentile capping shows why direct aggregation is unstable, but it does not by itself define a code-aware complexity metric.


\begin{tcolorbox}[breakable=true]
\noindent\textbf{Issue 2: Perplexity is heavy-tailed and spike-dominated across code structures.}
PPL is structured by code category, but remains highly skewed within categories and can be dominated by a small number of extreme tokens. These spikes~may reflect meaningful localized uncertainty, but they may also arise from rare tokens. Thus, token PPL should not be directly aggregated into a snippet-level understandability score. A viable PPL-based metric must control upper-tail domination and distinguish code-relevant surprise from artifact-driven rarity.

\end{tcolorbox}

\subsection{RQ3: How does human-label reliability affect perplexity alignment with behavioral difficulty?}

RQ2 shows that token-level PPL is noisy on the model side: it is heavy-tailed, spike-dominated, and sensitive to artifact-driven rarity. RQ3 examines a complementary source of noise: the human target used to evaluate alignment. Code understandability is not an absolute property of a snippet alone. The same code may be easy for a human but difficult for another, depending on 
their experience, familiarity, domain knowledge, and reasoning strategy. Thus, weak PPL-understandability alignment may arise not only from noisy model scores, but also from ambiguous or low-consensus human labels.

\subsubsection{\bf \em Empirical Settings}

Unless otherwise stated, RQ3 reuses the $\alpha$ dataset~\cite{scalabrino2021automatically} and the Qwen2.5-Coder-7B-Instruct teacher-forced token-PPL scores from RQ2. For this dataset, we use the released comprehension-question responses and define behavioral difficulty as
$\mathrm{Diff}(P) = 1 - \mathrm{CorrectnessRate}(P)$,
where larger values indicate greater human difficulty. We also evaluate the $\beta$ dataset~\cite{nguyen2026effectcodeobfuscationhuman}, which contains output-prediction tasks. For both datasets, the all-item setting keeps every item with a behavioral label.

To study label reliability, we define a {\bf \em consensus-filtered} setting. Items with correctness rate at or below $0.4$ are treated as mostly incorrect, items with correctness rate at or above $0.6$ are treated as mostly correct, and items in the middle band $(0.4,0.6)$ are treated as mixed/no-consensus for the primary alignment analysis. We use the 0.4--0.6 interval as a symmetric no-consensus band around chance-like split responses; items outside this band have a clearer majority direction and are used as the primary consensus-filtered set.
We also performed the study with consensus thresholds from 0.1--0.9 to 0.45--0.55 (see full results in the replication package~\cite{ccp-artifact}).
Across these thresholds, the conclusion remains unchanged: different consensus filtering does not make naive PPL summaries consistently reliable.
For the $\alpha$~\cite{scalabrino2021automatically} and $\beta$~\cite{nguyen2026effectcodeobfuscationhuman} datasets, item-level correctness rates define the all-item and consensus-filtered label sets (Table~\ref{tab:rq2-consensus-counts}). The retained column reports the mostly incorrect/correct items kept in the consensus-filtered set.

\begin{table}[t]
\centering
\scriptsize
\caption{Human-response consensus split for RQ3. Mostly incorrect items have correctness rate $\leq0.4$, mostly correct items have correctness rate $\geq0.6$, and mixed/no-consensus items fall between these bounds. Retained items form the consensus-filtered label set. (RQ3)}
\label{tab:rq2-consensus-counts}
\vspace{-3pt}
\begin{tabular}{lrrrrr}
\toprule
Data & Items & Mostly inc. & Mixed & Mostly corr. & Retained \\
\midrule
$\alpha$~\cite{scalabrino2021automatically} & 50 & 16 & 25 & 9 & 25 (50\%) \\
$\beta$~\cite{nguyen2026effectcodeobfuscationhuman} & 20 & 8 & 8 & 4 & 12 (60\%) \\
\bottomrule
\end{tabular}
\end{table}

\begin{table}[t]
\centering
\scriptsize
\caption{PPL alignment with behavioral difficulty. Values are Spearman $\rho$ with $p$ in parentheses for percentile, median, and IQR-normalized summaries under all-item and consensus label sets. (RQ3)}

\label{tab:rq2-iqr-first-alignment}
\tabcolsep 2.5pt
\vspace{-3pt}
\begin{tabular}{llrrrrr}
\toprule
Data & Label set & P90 & P80 & P70 & Median & IQR \\
\midrule
$\alpha$~\cite{scalabrino2021automatically} & all & 0.115 (.435) & 0.060 (.684) & -0.050 (.738) & -0.019 (.897) & -0.060 (.686) \\
$\alpha$~\cite{scalabrino2021automatically} & consensus & 0.235 (.259) & 0.144 (.491) & 0.005 (.980) & 0.019 (.929) & 0.022 (.917) \\
$\beta$~\cite{nguyen2026effectcodeobfuscationhuman}  & all & 0.133 (.577) & 0.339 (.144) & 0.275 (.240) & 0.248 (.292) & -0.296 (.206) \\
$\beta$~\cite{nguyen2026effectcodeobfuscationhuman}  & consensus & -0.162 (.616) & 0.190 (.555) & 0.095 (.769) & 0.042 (.896) & -0.373 (.233) \\
\bottomrule
\end{tabular}
\end{table}


\subsubsection{\bf \em Human-Response Consensus Result}

Table~\ref{tab:rq2-consensus-counts} shows that human responses do not always provide a clear snippet-level understandability label. In both the $\alpha$~\cite{scalabrino2021automatically} and $\beta$~\cite{nguyen2026effectcodeobfuscationhuman} datasets, a substantial fraction of items falls into the middle band between $0.4$ and $0.6$ correctness: 25 out of 50 items in the former and 8 out of 20 items in the latter. These items are neither mostly incorrect nor mostly correct; instead, participants disagree about whether the same code is understandable. Thus, half of the $\alpha$~\cite{scalabrino2021automatically} items and 40\% of the $\beta$~\cite{nguyen2026effectcodeobfuscationhuman} items do not provide a clear behavioral target under this threshold.

The mixed/no-consensus counts are therefore not discarded because the snippets are invalid; they are held out because aggregate correctness does not give a stable direction for evaluating PPL alignment. This pattern suggests that understandability depends not only on the code itself, but also on human background, familiarity, and reasoning. Thus, treating all items as a single ranked ground truth can introduce label noise into PPL-alignment analysis.

\subsubsection{\bf \em Naive PPL Alignment with Behavioral Difficulty}


We investigate whether consensus filtering alone makes naive PPL aggregation align with behavioral difficulty. Table~\ref{tab:rq2-iqr-first-alignment} shows Spearman correlations between behavioral~difficulty and several PPL summaries. P90, P80, and P70 are within-snippet percentile summaries; Median is the snippet-level~token median. For IQR, we normalize token surprisal within~each snippet by subtracting the snippet median and dividing by their IQR, keep only positive deviations, and aggregate their~median.


Table~\ref{tab:rq2-iqr-first-alignment} reports the correlations between behavioral difficulty and naive PPL summaries under both all-item and consensus-filtered label settings. 
Because Table~\ref{tab:rq2-iqr-first-alignment} compares multiple aggregation choices across small all-item and consensus-filtered slices, we treat individual $p$-values as exploratory and interpret the pattern of effect directions and stability across settings rather than any single cell.
Overall, consensus filtering alone is not sufficient to make PPL a reliable proxy for code understandability. Across datasets and aggregation choices, the correlations remain small to moderate and statistically insignificant; the largest absolute value is $|\rho|$=0.373 for the $\beta$ dataset~\cite{nguyen2026effectcodeobfuscationhuman} consensus IQR diagnostic, with $p$=0.233. The patterns are also inconsistent across aggregations. For the $\alpha$ dataset~\cite{scalabrino2021automatically}, P90 becomes more positive after consensus filtering ($\rho$=0.115 to $\rho$=0.235), but P80, P70, Median, and IQR remain near zero. For the $\beta$ dataset~\cite{nguyen2026effectcodeobfuscationhuman}, several all-item correlations are positive, but the consensus-filtered correlations become weaker or even negative. Thus, label consensus is a useful diagnostic for clarifying the human-difficulty target, but it is not a~complete solution: cleaner human labels help interpret PPL alignment, while simple PPL aggregation remains unstable.

\begin{table}[t]
\centering
\scriptsize
\caption{Participant agreement in FRP-AoC by the number of AOI regions per snippet. Agreement is computed as item-level pairwise agreement over participants' fixations. (RQ3)}
\label{tab:rq2-frp-aoc-aoi-count-agreement}
\tabcolsep 2.5pt
\vspace{-6pt}
\begin{tabular}{lrrccc}
\toprule
\# of AOI & Snip. & \#pers/snippet & Agree. [95\% CI]/$p$ & Dom. AOI [95\% CI]/$p$ \\
\midrule
1 AOI & 42 & 8.91 & 0.979 [0.953, 0.996] /$<$.001 & 0.989 [0.976, 0.998]  $<$.001 \\
2 AOIs & 42 & 10.50 & 0.620 [0.552, 0.693] /$<$.001 & 0.727 [0.671, 0.785]  $<$.001 \\
3 AOIs & 60 & 10.28 & 0.493 [0.453, 0.535] /$<$.001 & 0.660 [0.626, 0.694]  $<$.001\\
\bottomrule
\end{tabular}
\end{table}

\begin{figure}[t]
    \centering
    \includegraphics[width=3.7in]{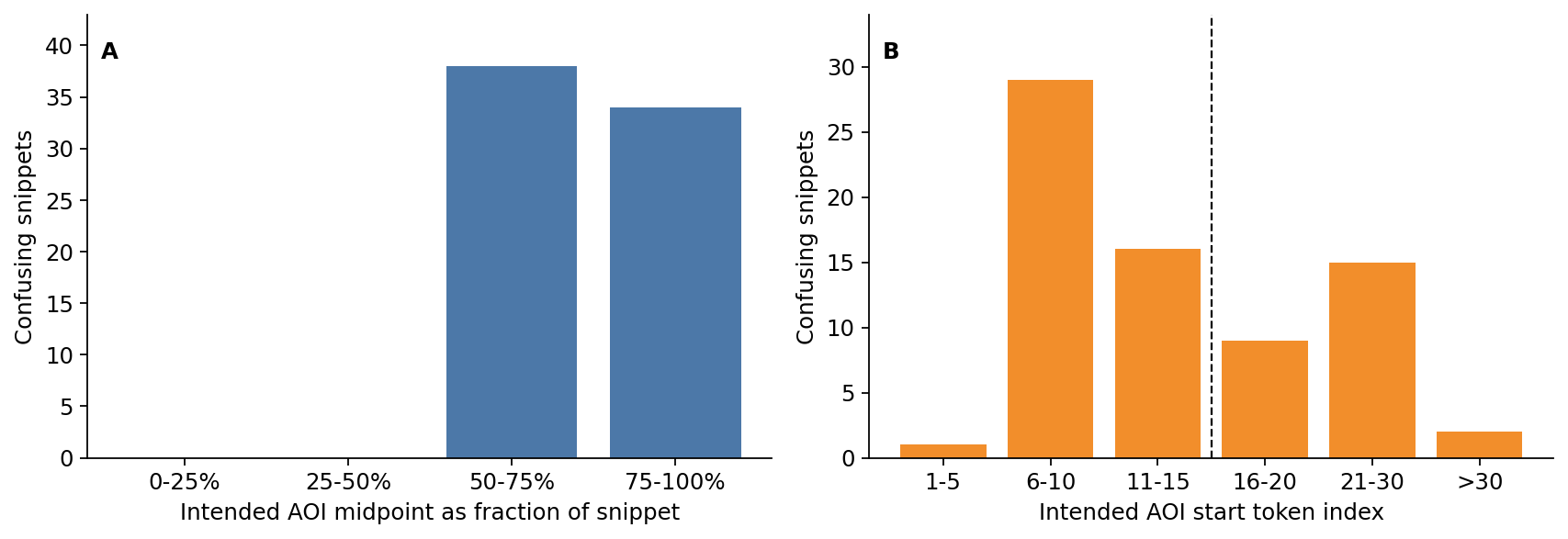}
    \vspace{-18pt}
    \caption{Position of intended AOI regions in the FRP-AoC dataset. The normalized and absolute-token views show that AOIs are late within short snippets but still occur within the first few Qwen tokens. (RQ3)}
    \label{fig:rq2-frp-aoi-position}
\end{figure}


\subsubsection{\bf \em Why FRP-AoC Shows Localized PPL--Confusion Alignment for Abdelsalam {\em et al.}~\cite{abdelsalam2025humansllmsprocessconfusing}}

Unlike the $\alpha$~\cite{scalabrino2021automatically} and $\beta$~\cite{nguyen2026effectcodeobfuscationhuman} datasets, whose difficulty labels are inferred from participant answers over entire snippets, FRP-AoC defines clean/confusing contrasts by design and associates the human response with intended areas of interest (AOIs). Thus, the primary evidence in this dataset is AOI-level alignment rather than whole-snippet analysis. In Abdelsalam {\em et al.}~\cite{abdelsalam2025humansllmsprocessconfusing}, FRP-AoC (Fixation-Related Potentials on Atoms of Confusion) denotes the fixation-related-potential dataset built around atoms of confusion. It provides a localized human-confusion signal: each snippet contains a designed area of interest corresponding to a confusing code construct or its clean counterpart, and human response is measured through EEG signals time-locked to participant fixations on that region.

This design avoids one major difficulty in our snippet-level datasets: the human target is not reconstructed from aggregate correctness over a full method. In the $\alpha$~\cite{scalabrino2021automatically} and $\beta$~\cite{nguyen2026effectcodeobfuscationhuman} datasets, a snippet may receive mixed responses because different participants understand different parts of the code, apply different reasoning strategies, or make different mistakes. FRP-AoC instead constrains the expected source of confusion before measurement: the confusing construct is known by design, and the EEG response is evaluated at the corresponding AOI. Therefore, the label is more localized and less dependent on inferring a single whole-snippet difficulty score from heterogeneous participant behavior.



Fig.~\ref{fig:rq2-frp-aoi-position} further shows that the intended AOIs occur inside short snippets. In normalized position, the AOIs are usually located in the latter half of the snippet; in absolute token position, they still occur within a small number of tokens because the snippets themselves are short. This matters because the FRP-AoC analysis compares PPL and human response at a localized region with limited surrounding code, rather than aggregating PPL over an entire real-world method.

Table~\ref{tab:rq2-frp-aoc-aoi-count-agreement} provides an additional check on whether participants concentrate on the designed regions. Agreement is highest for snippets with a single AOI and decreases when multiple AOIs compete for attention. This pattern is expected: when there is only one designed confusing region, participants' fixations are more consistently assigned to the same target; when several AOIs are present, attention can be distributed across multiple plausible regions. Importantly, this agreement measures fixation-assignment concentration, not correctness-style consensus. Thus, FRP-AoC is not equivalent to consensus-filtered behavioral datasets, but it reduces label ambiguity because the human signal is tied to designed code regions rather than to whole-snippet answer correctness.



In brief, perplexity aligns with human confusion in FRP-AoC dataset in Abdelsalam {\em et al.}'s study because:

1. FRP-AoC has localized AOI-level labels, not whole-snippet behavioral labels.

2. It avoids reconstructing difficulty/understandability from aggregate correctness.

3. AOIs occur in short snippets and are evaluated locally.

4. Participant agreement is high when there is one AOI and lower when multiple AOIs compete.

Therefore, the FRP-AoC result supports the value of PPL as a localized cognitive signal, while our results show that additional care is required before that signal can be lifted to snippet-level code understandability.

\begin{tcolorbox} [breakable=true]
\noindent\textbf{Issue 3: Human-label disagreement weakens snippet-level alignment.}
Human understandability is not determined by code alone; it also depends on developer background, API familiarity, reasoning strategy, and task-specific factors. In our $\alpha$~\cite{scalabrino2021automatically} and $\beta$~\cite{nguyen2026effectcodeobfuscationhuman} datasets, many snippets fall into a mixed-response band, indicating that participants do not consistently agree whether the same code is easy or difficult. Consensus filtering creates a clearer behavioral target, but it does not make naive PPL aggregation reliable. Thus, human-label reliability must be considered when evaluating PPL-based understandability measures, but cleaner labels alone are not sufficient.

\end{tcolorbox}

\subsection{RQ4: How model-dependent are PPL distributions and their alignment with human difficulty?}
\label{RQ4}


\begin{figure*}[h]
    \centering
    \includegraphics[width=0.75\textwidth]{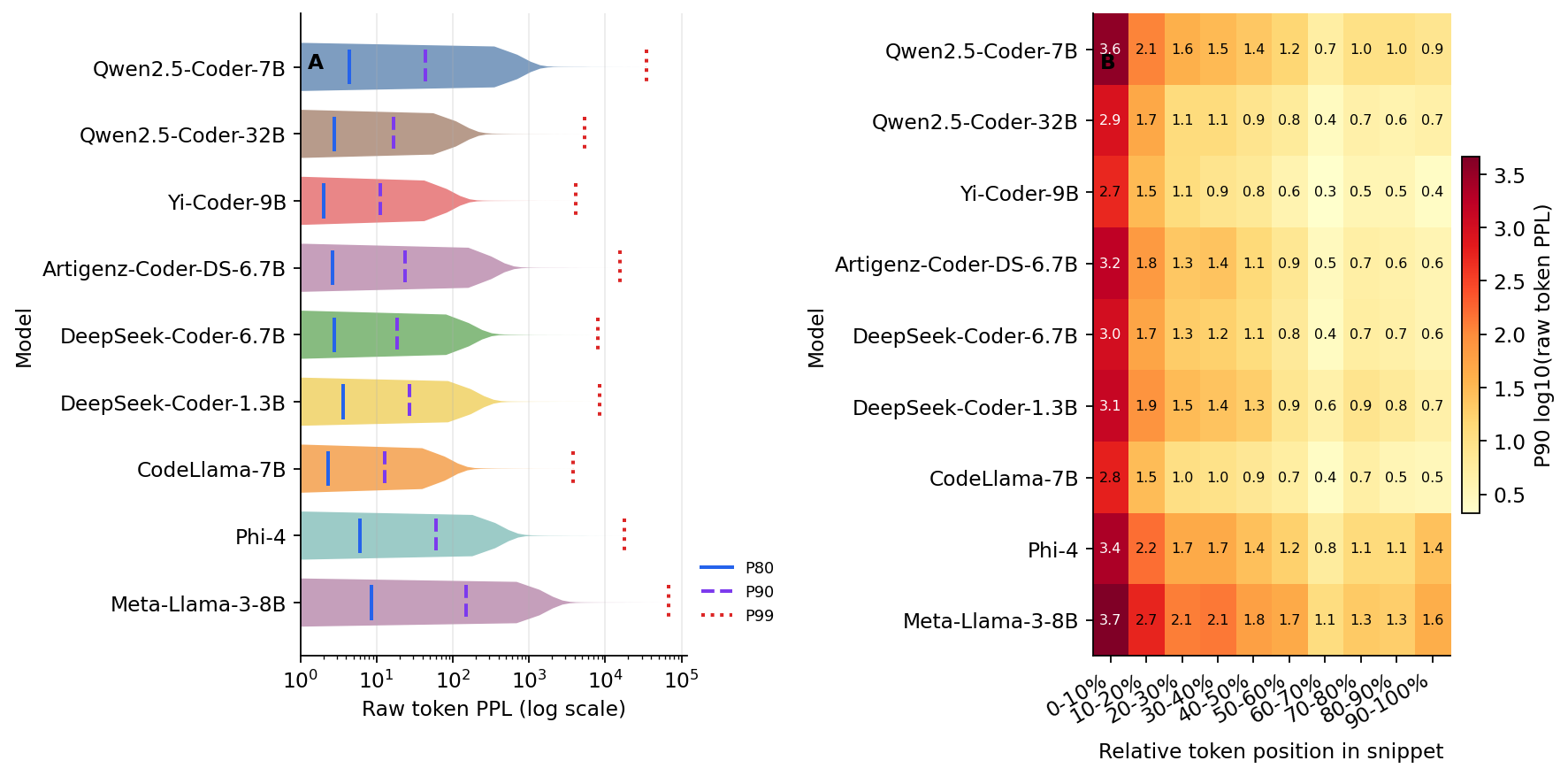}
    \vspace{-9pt}
    \caption{PPL distributions over the same 50 snippets from the $\alpha$ dataset~\cite{scalabrino2021automatically} across model/tokenizer choices. The overall and position-binned views show that PPL scale and upper-tail behavior are model-dependent before any human-label alignment analysis. (RQ4)}
    \label{fig:rq4-model-tokenizer-perplexity-variation}
\end{figure*}

Although PPL is mathematically model-dependent by definition, this dependence becomes problematic when PPL is treated as a reusable code-level metric: the same code can receive different tail behavior, position profiles, and difficulty correlations depending on the chosen scorer. 

\subsubsection{\bf \em Empirical Settings}


RQ4 uses two diagnostics. The first is a {\bf \em distributional diagnostic}: we score the same snippets from the $\alpha$ dataset~\cite{scalabrino2021automatically} with multiple models and compare non-whitespace token-level PPL distributions without using human labels. This asks whether PPL is on a comparable scale across models. The second is an {\bf \em alignment diagnostic}: we compute the same PPL summaries for each model and correlate them with behavioral difficulty on the consensus-filtered $\alpha$~\cite{scalabrino2021automatically} and $\beta$~\cite{nguyen2026effectcodeobfuscationhuman} datasets. This asks whether changing the scoring model makes naive PPL aggregation more reliable.

For each model and dataset item, we compute item-level P80 and Median PPL over that item's token scores, and then correlate those item-level summaries with behavioral difficulty using Spearman $\rho$ on the consensus label set. Table~\ref{tab:rq4-model-sensitivity} uses the same available model set as Fig.~\ref{fig:rq4-model-tokenizer-perplexity-variation}.
Llama-3.2-3B is listed only to document that it was unavailable since its repo was gated in the local environment. 
We report the full P90, P80, P70, Median, and IQR sweep in the replication package~\cite{ccp-artifact}.

\subsubsection{\bf \em Empirical Results}

Fig.~\ref{fig:rq4-model-tokenizer-perplexity-variation} is not an alignment result; it is a measurement-stability one.
We plot these values on the $\log_{10}$ PPL scale and cap the displayed violin density at each model's P99 value so that the dense body and high-tail percentiles remain visible in the same panel. This cap is used only for visualization; all percentile markers are computed from each model's original, uncapped token-PPL distribution.
The central token-level~values may look similarly low across models, but the upper tail differs substantially. This matters because RQ2 already showed that snippet-level aggregation is sensitive to upper-tail spikes. If the same code produces different high-tail behavior under different tokenizers and model distributions, then a fixed PPL threshold or percentile summary cannot be assumed to measure the same code property across models.

The position-binned view in Fig.~\ref{fig:rq4-model-tokenizer-perplexity-variation} adds a second source of model dependence. All models show larger upper-tail PPL near the beginning of snippets, where less preceding context is available, but the size and persistence of this early-position effect differ by model. Thus, model choice changes not only the magnitude of PPL but also where the high-PPL tail appears inside the snippet. A code unit can therefore appear more or less surprising because of model/tokenizer behavior, not only because of the code's human difficulty.
This position effect reinforces the RQ2 concern: some high-PPL values may reflect context availability in the scoring setup rather than code regions that are cognitively difficult for humans.

Table~\ref{tab:rq4-model-sensitivity} shows that no model produces a stable positive alignment across both datasets and both aggregation choices. The $\alpha$~\cite{scalabrino2021automatically} correlations are sometimes moderately positive but not significant; the $\beta$~\cite{nguyen2026effectcodeobfuscationhuman} correlations are weak or negative, with Median negative for every available model. Thus, model choice can change the apparent direction and strength of PPL alignment, but it does not make naive PPL aggregation reliable. The significant negative DeepSeek-Coder cell should be interpreted cautiously because the $\beta$~\cite{nguyen2026effectcodeobfuscationhuman} consensus slice is small; nevertheless, it illustrates that model choice can even reverse the expected direction of alignment.
Because Table~\ref{tab:rq4-model-sensitivity} compares multiple models, datasets, and aggregation choices on small consensus-filtered samples, we treat individual $p$-values as exploratory and interpret the cross-model pattern rather than any single significant cell.

%

\begin{table}
\caption{Model-set sensitivity of PPL alignment on consensus label sets. Values are Spearman $\rho$ with $p$ in parentheses. (RQ4)}
\label{tab:rq4-model-sensitivity}
\setlength{\tabcolsep}{2.5pt}
\vspace{-6pt}
\resizebox{\columnwidth}{!}{%
\begin{tabular}{lcccc}
\toprule
Model & $\alpha$ P80 & $\alpha$ Med. & $\beta$ P80 & $\beta$ Med. \\
\midrule
Qwen-7B & .190 (.362) & .262 (.206) & .144 (.655) & -.063 (.845) \\
Qwen-32B & .153 (.467) & .199 (.341) & -.018 (.957) & -.274 (.388) \\
Yi-9B & .382 (.059) & .237 (.253) & -.070 (.828) & -.313 (.322) \\
Artigenz-6.7B & .381 (.060) & .282 (.173) & -.267 (.401) & -.548 (.065) \\
DeepSeek-6.7B & .353 (.084) & .252 (.224) & -.172 (.592) & -.640 (.025) \\
Llama-3.2-3B & \multicolumn{4}{c}{Unavailable (gated repository)} \\
DeepSeek-1.3B & .317 (.122) & .250 (.228) & -.172 (.592) & -.141 (.663) \\
CodeLlama-7B & .154 (.461) & .124 (.556) & -.190 (.555) & -.211 (.511) \\
Phi-4 & .318 (.122) & .295 (.152) & -.105 (.744) & -.214 (.503) \\
Meta-Llama-3-8B & .237 (.255) & .245 (.239) & -.074 (.820) & -.330 (.294) \\
\bottomrule
\end{tabular}
}
\end{table}

In brief, RQ2 shows that token PPL is heavy-tailed within code structures. RQ3 shows that consensus filtering clarifies the human target but does not make naive PPL aggregation reliably align with behavior. RQ4 adds that the PPL signal itself is model- and tokenizer-dependent. Therefore, poor alignment is not simply a failure of one model or one cutoff. These results do not support treating a convenient LLM plus a simple percentile or median aggregation rule as a reliable snippet-level understandability metric.

\begin{tcolorbox}
{\bf Issue 4}: {\bf \em Model dependence}. PPL scale, upper-tail behavior, position profile, and behavioral alignment all change with the model/tokenizer used to score the same snippets. Simple PPL aggregation is therefore not a model-invariant understandability metric.
\end{tcolorbox}

\subsection{Threats to Validity}
\label{sec:threats}

\subsubsection{Generalizability}

Our results may not generalize to all languages, domains, developer populations, or comprehension tasks. Although our datasets include Java/Android methods, Python/JavaScript output-prediction tasks, Java review changes, and FRP-AoC snippets, they are limited in scale and scope. Thus, our findings should be viewed as diagnostic evidence about snippet-level PPL aggregation, not as a universal conclusion about code understandability.

\subsubsection{Construct Validity}

Our labels are proxies for~underst\-andability and confusion, not complete measurements.~Com\-prehension-question correctness, output-prediction accuracy, review changes, and FRP-AoC signals capture different aspects of comprehension. Human disagreement also remains a concern: consensus filtering clarifies the behavioral target, but reduces sample size and cannot remove all label~ambiguity.

\subsubsection{Internal Validity}

Our conclusions depend on choices in PPL computation, token alignment, AST categorization, aggregation, and statistical analysis. PPL can be affected by tokenization, rare identifiers, literals, formatting, API-specific names, and left-context availability. We mitigate these risks with multiple datasets, aggregation diagnostics, consensus-aware analyses, and model-sensitivity checks.

\section{Related Work}
\label{sec:literatureAndDiscussion}




Classical metrics such as cyclomatic complexity~\cite{mccabe1976complexity}, nesting depth, Halstead measures~\cite{halstead1977elements}, and token/count-based features capture structural or lexical properties of code, but prior studies have shown that such metrics often provide limited alignment with human understandability~\cite{scalabrino2021automatically}.

Human-centered studies provide complementary evidence on code difficulty.
Sharif and Maletic~\cite{sharif2010camelcase} used eye tracking to study identifier naming styles, while Busjahn \emph{et al.}~\cite{busjahn2015eye} showed that code reading does not strictly follow the linear order of natural-language text. Peitek, Siegmund, and Apel~\cite{peitek2020readingorder} further studied what drives programmers' reading order and found that source-code linearity, expertise, and comprehension strategy all influence gaze behavior. Neurophysiological studies complement these findings by relating program comprehension to cognitive load and neural efficiency~\cite{siegmund2017neural,peitek2023efficacy}. Together, these studies suggest that human difficulty is reflected in localized attention and cognitive effort, not merely in whole-program structural metrics.

Recent work also uses human attention data to improve neural models of code. Bansal, Sharif, and McMillan~\cite{bansal2023humanattention} model human attention from eye movements and use it to improve neural source-code summarization. EyeTrans~\cite{zhang2024eyetrans} integrates human attention into Transformer-based code summarization, showing that gaze-derived signals can guide model attention toward code regions that humans find salient. More recent work further explores training code models to mimic human visual attention~\cite{zhang2025eyemulator}. These studies indicate that human behavioral signals can provide useful supervision for code models, but they primarily use gaze as a training or alignment signal rather than as a direct metric of code difficulty.


Paltenghi and Pradel~\cite{paltenghi2021thinking} compare the attention weights of neural models of code with human visual attention during code summarization, asking whether models attend to the same tokens as developers. Kou \emph{et al.}~\cite{kou2024attention} extend this to LLM-based code generation and report systematic misalignment between LLM attention and human attention. These studies are closely related to our goal of connecting model behavior with human code understanding, but they focus on attention alignment.

\section{Conclusion and Contribution}
\subsubsection*{\bf \em Implications for Future Research}



Our study contributes to the growing effort to study whether signals from LLM models can serve as computational proxies for code understandability. 

\textbf{Contribution 1: A systematic assessment of perplexity as a code-understandability proxy.}
Our first contribution is to provide a systematic investigation of whether language-model perplexity can model human code understandability at the code-snippet level. Prior work~\cite{abdelsalam2025humansllmsprocessconfusing} suggests that model surprisal may align with human confusion at localized code regions.
Our findings show that the hypothesis is not supported in its naive snippet-level form. Simple snippet-level aggregations, such as whole-snippet perplexity or top-token averages, are unstable and often weakly aligned with human understandability labels. This result is important as it challenges an intuitive but overly broad assumption: that code which is surprising to a language model is necessarily difficult for humans. Instead, our results indicate that model surprise must be interpreted with respect to code structure, token position, and semantic role.

\textbf{Contribution 2: Evidence that perplexity is confounded by modeling and measurement artifacts.} Our second contribution is to identify several confounding factors that limit the direct use of perplexity for code understandability. First, autoregressive models exhibit strong position sensitivity: early tokens often receive high surprisal because little context is available, even when those tokens are not cognitively difficult for humans. Second, perplexity is model- and tokenizer-dependent, making values difficult to compare across models or even across code snippets tokenized differently. Third, high-surprisal tokens may correspond to syntactic artifacts, rare literals, separators, formatting choices, or library-specific identifiers, rather than genuine cognitive difficulty. Fourth, the human target itself can be noisy: many snippets receive mixed participant responses, and consensus filtering clarifies the behavioral difficulty label but does not make naive perplexity aggregation reliable. These findings imply that perplexity should be treated as a noisy signal of model expectation, and that its alignment with human understandability must be evaluated with consensus-aware human labels rather than assumed from PPL values alone.


\textbf{Contribution 3: A shift from snippet-level perplexity to localized surprisal analysis.}
Another contribution is~to~clarify why the value of perplexity is more likely to stay in localized analysis than in whole-snippet scoring. Human difficulty is often concentrated in specific regions: a confusing branch~condition, a misleading identifier, an unusual API, a non-obvious data-flow dependency, or a boundary between conceptual steps. Token-level surprisal can help identify such regions, but only if the aggregation method preserves locality. This suggests that future metrics should move beyond treating a snippet as a flat sequence of tokens. Instead, perplexity should be aggregated over semantic units such as expressions, statements, control predicates, data-flow slices, or areas of interest. Such localized measures are more likely to capture the parts of code where humans experience cognitive~friction.

\textbf{Contribution 4: Implications for designing model-based understandability metrics.}
Our results suggest several principles for future research on model-based code understandability. First, snippet-level code understandability should not~be modeled by a global perplexity score. A snippet may contain a few highly surprising tokens that do not affect human comprehension, or it may contain low-surprisal code whose behavior is still difficult to reason about due to subtle dependencies. Thus, perplexity-based metrics should be used as elements of a richer framework rather than as standalone complexity scores.

Second, future metrics should separate semantic surprise from incidental token surprise. A rare variable name, literal, separator, or formatting artifact should not contribute in the same way as a surprising control-flow condition, unexpected API, or non-obvious data dependency. Third, metrics should correct for positional effects introduced by autoregressive prediction and should be calibrated within model and tokenizer families rather than compared using perplexity values. Fourth, metrics should aggregate surprisal over code-relevant units, rather than over flat code sequences.

These findings point to several directions for future work. One direction is to develop AST-aware, statement-aware, or slice-aware surprisal metrics that normalize for model and position effects while filtering out non-semantic artifacts. Another is to combine model surprisal with human behavioral data, such as response accuracy, response time, eye fixation, or neurophysiological measures, to build more faithful hybrid metrics. Finally, future work should evaluate whether perplexity-based localization can support practical tools, such as highlighting confusing code regions, recommending simplifications, or guiding code-review attention. Under this view, perplexity is not a direct measure of understandability by itself; it is a lens for identifying where a model's expectations are violated, which can then be related to human cognitive effort.

Our results provide a cautionary but constructive message. Perplexity is a promising signal for modeling code understandability, but only when used carefully. Whole-snippet perplexity is too coarse and confounded to serve as a reliable standalone metric. However, localized, normalized, and semantically filtered surprisal has the potential to become a useful component of future human-centered code complexity measures. 

\vspace{1pt}
{\bf Data Availability Statement.} To facilitate future research, we make our replication package publicly available~\cite{ccp-artifact}.

\balance
\bibliographystyle{IEEEtran}
\bibliography{ref.bib,references}

\pagebreak
\end{document}